# Blowing Polar Skyrmion Bubbles in Oxide Superlattices

Zijian Hong and Long-Qing Chen

Department of Materials Science and Engineering, The Pennsylvania State University, University Park, PA 16802, USA

Particle-like topological structures such as skyrmions and vortices have garnered ever-increasing interests due to the rich physical insights and potential broad applications. Here we discover the reversible switching between polar skyrmion bubbles and ordered vortex arrays in ferroelectric superlattices under an electric field, reminiscent of the Plateau-Raleigh instability in fluid mechanics. Electric field phase diagram is constructed, showing wide stability window for the observed polar skyrmions. This study is a demonstration for the computational design of ferroelectric topological structures and field-induced topological phase transitions.

______________________________

Corresponding authors:

\* Dr. Z. Hong, Email: zxh121@psu.edu

\# Dr. L. -Q. Chen, Email: lqc3@psu.edu

A plethora of particle-like topological defects such as skyrmions with whirl-like no-trivial topological structures [1-5] and vortices with continuous rotating order parameter vectors [6-8] have been widely investigated in the ferromagnetic systems for decades. It is now well accepted that the antisymmetric chiral interactions such as Dzyaloshinskii-Moriya (DM) interaction [9, 10] in the non-centrosymmetric ferromagnetic systems could give rise to the rotating spontaneous magnetization, benefiting from the spin-orbit coupling. Topological polar structures on the other hand are considered rare in the ferroelectric systems because such chiral interactions are absent since the fundamental origin of the ferroelectrics is different from the ferromagnetics [11, 12].

The recent progresses in computational tools have enabled the predictive modeling of mesoscale polar topological transitions. One particular example is the state-of-art phase-field simulation [13], which is not only capable of simulating mesoscale microstructure evolution, but also allows predictions that can be validated by the experiments. For instance, a recent phase-field study has predicted the whole range of periodicity phase diagram for $(PbTiO_3)_n/(SrTiO_3)_n$ (PTO/STO in short) superlattice on a $DyScO_3$ substrate and in particular polar vortex lattice at intermediate periodicities due to the complex intimate energy competitions, which is beautifully confirmed by experimental observations [14, 15].

The exciting discovery of polar vortex lattice has brought one key question as how to manipulate the ferroelectric topological structures via external stimuli. To date, the vortex switching and annihilation dynamics has been extensively investigated both experimentally and theoretically (e.g., micromagnetic simulations [16], phase-field simulations [17]) in ferromagnetics [18, 19] and superconductors [20]. Some intriguing phenomena have been discovered, for example vortex-antivortex annihilation in ferromagnetics could not only reverse the polarity of the vortex core but also induce a burst of spin waves [18]. In the ferroelectric systems, the switching of the curl of the polar vortex has been studied in low dimensional ferroelectrics [21-23]. Here we predict the field-driven topological phase transitions between polar vortices and skyrmions.

The switching of a polar vortex lattice under an electric field is studied via phase-field simulations (details in Supplementary Information). The equilibrium polar vortex lattice is constructed for a $(PTO)_{16}/(STO)_{16}$ superlattice. As shown in Figure 1(a), the polarization vectors rotate continuously inside the PTO layers, which agrees well with the structure from both theoretical calculations and experiment observations [14, 15]. The neighboring vortices show the opposite vorticity (indicated by the blue and red regions, as calculated by $\nabla \times \vec{P}$), representing the clockwise-counter clockwise geometric arrangement of the vortices. While the planar view image cutting from the top of the PTO layer indicates the formation of long vortex lines (Figure 1b), with one vortex line much thicker than the neighboring one with opposite in-plane polarization directions. The distance of the vortex cores in neighboring vortices is roughly 5-6 nm that scales with the PTO spacing of 16 unit cells [14, 15 and 24]. The three-Dimensional (3D) structure of the polarization inside one PTO layer is further plotted in Figure 1(c), where the long vortex tubes form perpendicular to the vortex plane. Meanwhile, this structure is highly asymmetric, forming an up-down zigzag configuration [24]. The out-of-plane field is then applied through a top capacitor electrode. The profile of time-dependent applied bias is given in Figure 1(d), which increases 1.3 V every 400 time steps until reaching a maximum of 13 V, then decreases at the same rate to -13 V, and eventually recovers to 0 V again to form a complete switching circle.

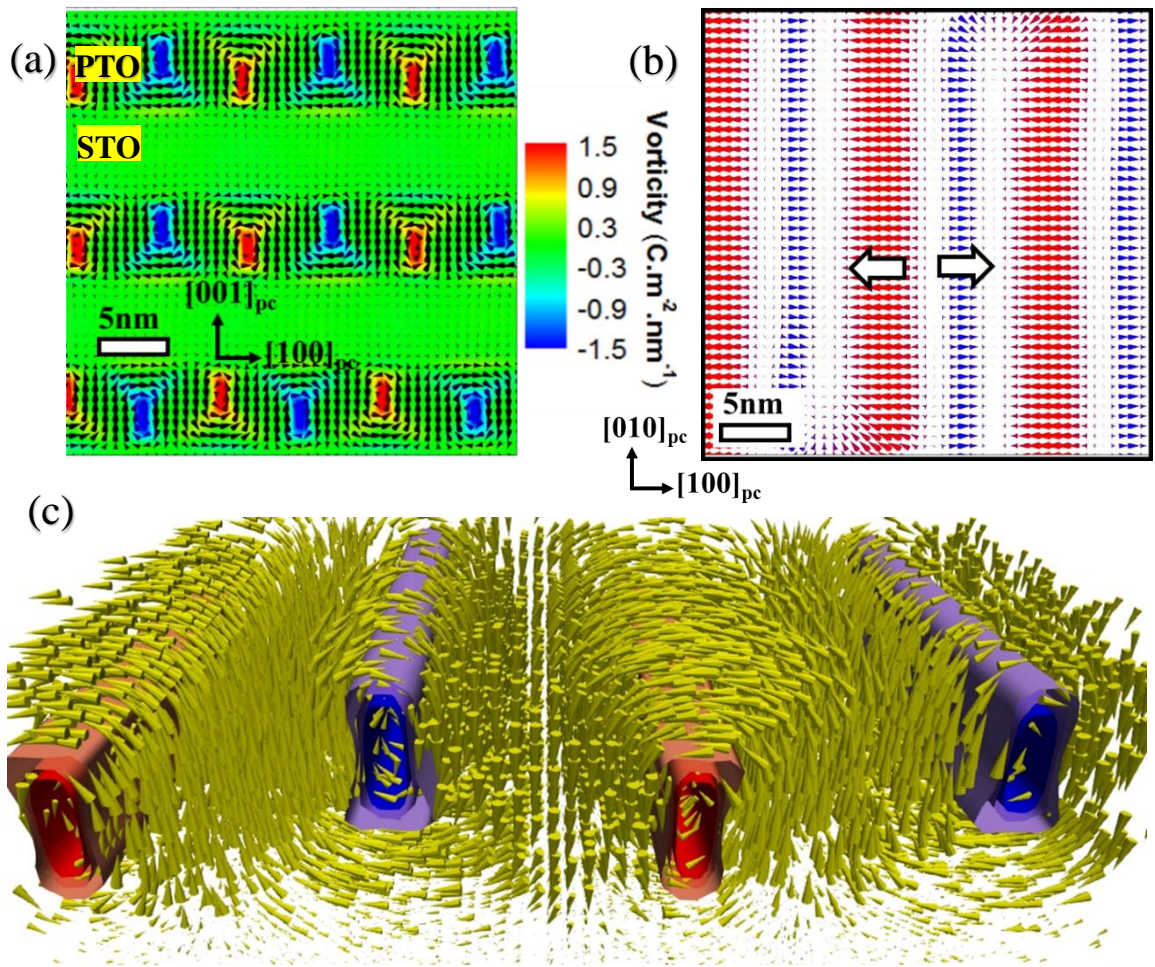

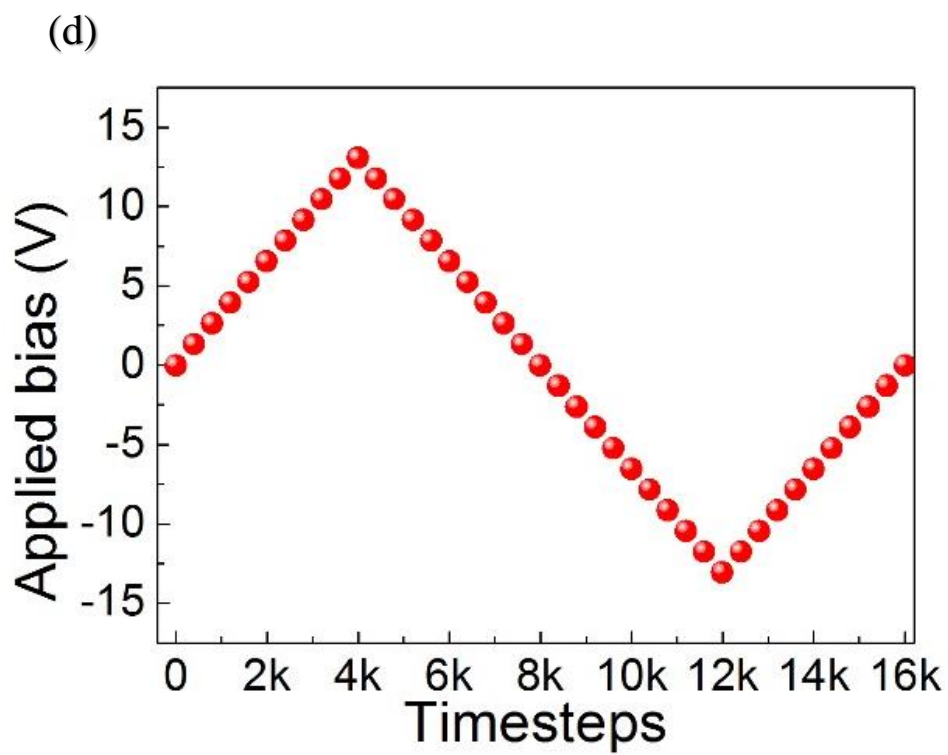

**Fig. 1. Equilibrium polar vortex pattern in the (PTO)$_{16}$/(STO)$_{16}$ superlattice assumed to be on a DSO (110)$_o$ substrate without an external electric field and the applied bias profile** (a) Cross-section view showing the vector plot of polar vortex lattice overlay on the spatial distribution of the vorticity. It is shown that the neighboring vortices have the opposite rotation directions, characterized by the alternating sign of vorticity; (b) Planar view showing the asymmetric vortex lines where the blue (polarization pointing right) and red regions (polarization pointing left) are not equal size; (c) 3-dimensional vector plot of the polarization in one PTO layer, blue and red tubes show the clockwise and counter-clockwise vortex cores with opposite vorticities; (d) Schematics of the applied bias in one switching cycle, the maximum bias is 13 V.

The cross-section view of the entire switching circle is shown in Figure 2(a)-(h). With the application of a positive bias (negative field), two neighboring vortices with opposite curls tend to move closer to form a close-pair structure, while the neighboring pairs are moving against each other, (Figure 2b). The lateral movement of the vortex cores could reduce the area with upward polarization, thus reducing the electrostatic energy of the whole simulation system. This switching process demonstrates that the polar vortex lattice exhibit "dipole spring"-like behavior [25]. Increasing the applied electric field will lead to the destruction of vortex arrays as soon as two vortex cores reach the same lateral position (Figure 2c). This process produces new *a*-domains, with the decrease in vorticity due to the annihilation of positive and negative vorticity regions. Consequently, at even higher biases, the full destruction of polar vortex lattice gives rise to regular *a*/*c*-twin domains.

Upon the gradual removal of applied field, the reversible back-switching takes place. The *a*/*c*

domain wall first decomposes into a pair of two vortices with opposite polarization vorticities. The two vortices then move laterally against each other to reduce the net polarization as to decrease the huge depolarization field. When applied field is removed, the ordered vortex array structure is restored, and zigzag vortex lattice pattern appears again (Figure 2d).

A negative bias is gradually applied then. It is observed that the lateral motion direction of vortices is reversed (Figure 2e) as compared to Figure 2(b) with a positive bias. This could be understood since the preferred polarization direction flips with the change of direction of the applied field. An even higher bias led to the destruction of the vortex lattice again, forming *a*/*c*-twin domain structure (Figure 2f). The orientation of the 90° domain wall switches as compared to Figure 2(c) with a positive potential. As the applied bias is further increased, the bottoms of *a* domains shrink to needle-like configurations before they are eventually switched to *c* domains. Consequently, the removal of the applied field leads to the full recovery of the vortex structure again. More details are shown in the supplementary movie S1.

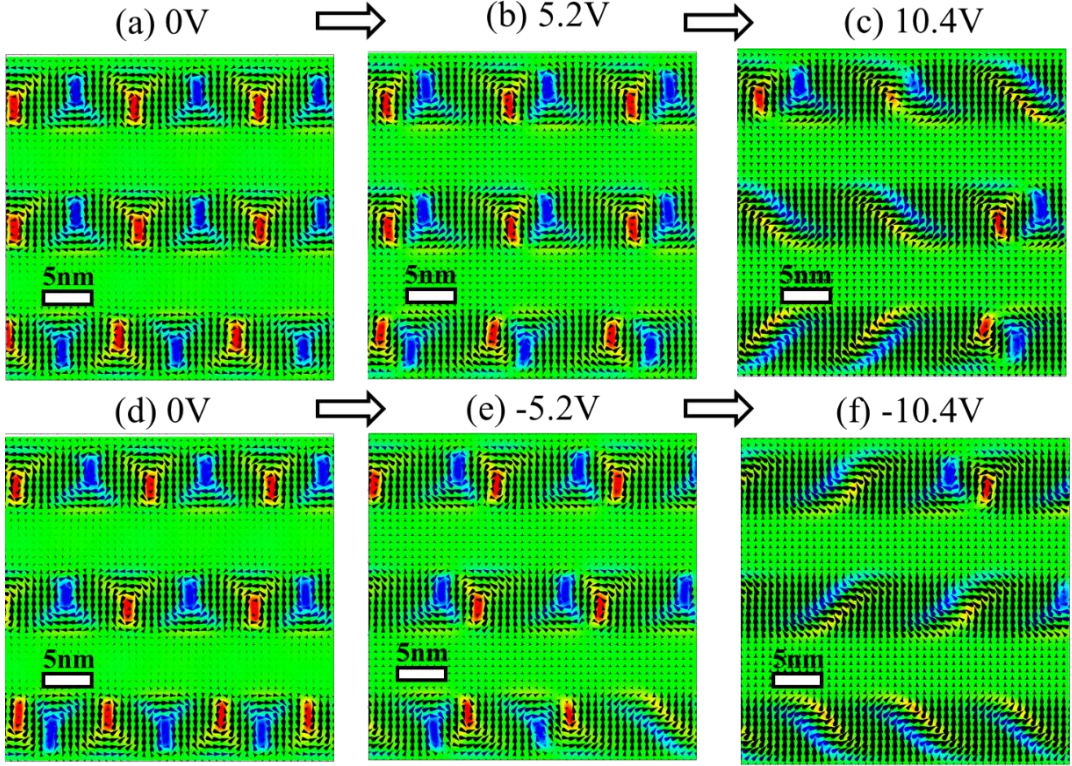

**Fig. 2. Cross-section view of the polarization pattern evolution under a small to large applied bias.** (a)-(c), under applied bias of 0, 5.2 and 10.4 V, respectively. (d)-(f), under applied bias of 0, -5.2 and -10.4V, respectively. The reversible transition of vortex pairs to the *a*/*c* twin structure is demonstrated. Red and blue illustrate the positive (counter-clockwise) and negative vorticities (clockwise), respectively.

Previously, it is demonstrated through integrated micromagnetic simulations and experimental observations that switching of ferromagnetic nano-domains with wall junctions could induce a ferromagnetic vortex structure, resulting in a dimensionality crossover from 2-D domain wall to 1-D vortex [26]. In the current simulation, we found a similar dimensionality crossover in the polar vortex system where 1-D vortices can be reversibly switch to 2-D domain walls.

The planar view vector plot is shown in Figure 3(a)-(f), corresponding to the structure of top PTO layer in Figure 2(a)-(f). The initial structure in Figure 3(a) indicate the formation of long vortex lines. Then, with the increase of applied bias, the size of both vortex lines shrinks, (see Figure 3b), which is due to the switching of the *a*-like regions to *c*-like regions under an out-of-plane field. Simultaneously, the two vortex lines move towards each other to further reduce the area with opposite polarization directions. With higher bias (Figure 3c), the thinner vortex lines "melt", forming "bubble"-like polar pattern. This is in analogy to the Rayleigh-Plateau instability in fluid mechanics, where the lines of waterfall break into small water droplets to minimize the surface area. Previously, Scott *et al.* have discovered the Richtmyer–Meshkov type instabilities in $Pb_5Ge_3O_{11}$, where the application of an external electric field leads to the emission of a bubble-like ferroelectric domains due to the curving of the domain walls [27, 28]. It is proposed that for some ferroelectric materials undergoing non-equilibrium processes, the motion of the ferroelastic domain walls can be controlled by fluid mechanics-like dynamics. Here we show another type of hydrodynamic instability in the ferroelectric/ferroelastic system under non-equilibrium conditions.

Consequently, the "bubble" size shrinks due to the further reduction of the area with upward polarization. This is primary realized by the shrinking of the "bubble" length along the vortex lines. Eventually, the "bubbles" burst at large applied biases, forming stripe *a/c*-twin domains. When the bias is gradually removed, the "bubble" forms again and expands along the *a*-domain stripe direction until the neighboring bubbles emerge to form vortex lines (Figure 3d). The distance between the neighboring vortex lines as well as the lengths of the vortex lines decrease, due to the switching of the downward polarizations, as indicated in Figure 3(e). The "bubble"-like domain forms again with a higher bias (Figure 3f), but the location of the bubble is flipped as compared to Figure 3(c). And the further movement of the bubble along the stripe leads to the shrinking of the bubble again and eventually disappears. More detailed dynamic process is shown in Supplementary movie S2.

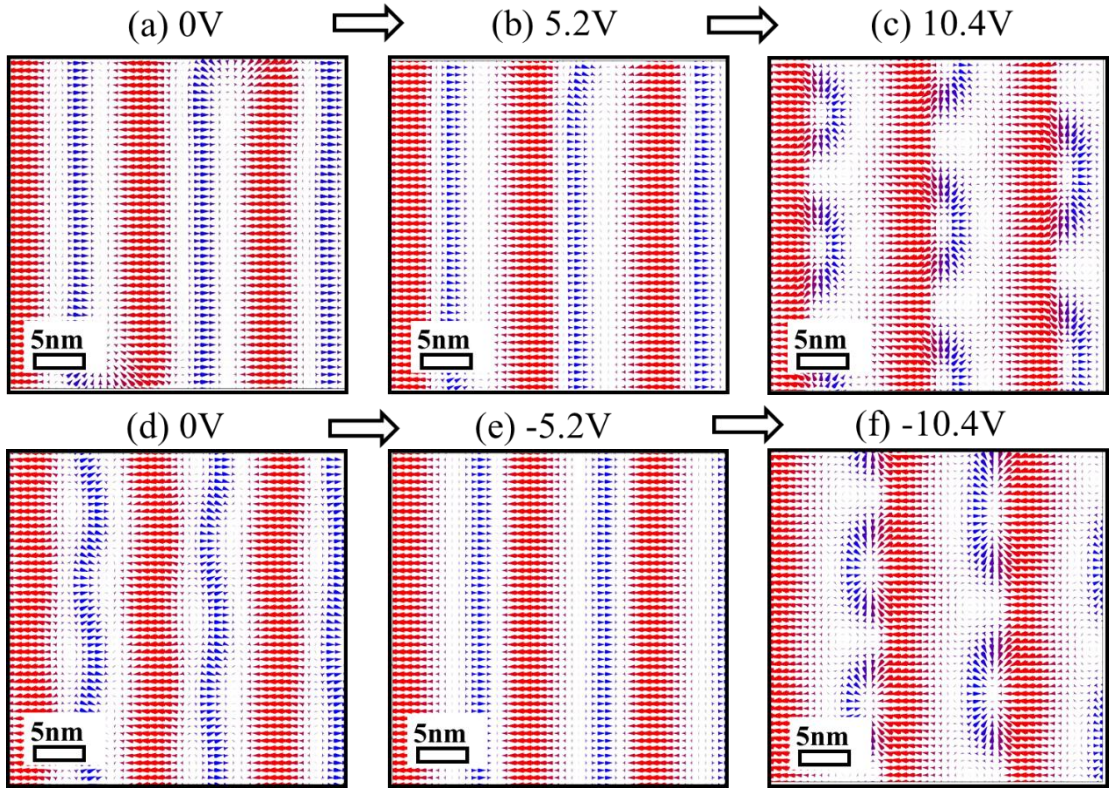

**Fig. 3. Planar view of the polarization pattern evolution under small to large applied bias.** (a)-(c), under applied bias of 0, 5.2 and 9.1 V, respectively. (d)-(f), under applied bias of 0, -5.2 and -9.1 V, respectively. The melting of the vortex line into "bubble" structure is shown.

In order to have a deeper understanding of the bubble structure, we plot the 3-D vector structure under applied bias of 9.1 V, shown in Figure 4(a). It can be clearly seen that the polarization rotates from upward inside the bubble core to downward through the in-plane direction, which is similar to the "Néel-type skyrmion" structure in ferromagnetics [29]. In the ferromagnetics, the ultrafast reversible transition between a domain wall and skyrmion via the application of electric current has been proposed through micromagnetic simulations [30]. Very recently, the "blowing" of magnetic skyrmion bubbles from stripe domains has been experimentally observed with careful design of the device geometry [31].

To further investigate the field-induced phase transitions, the electric field phase diagram is consequently constructed (Figure 4b). In the relatively low field region (<500 kV/cm), it is demonstrated that the vortex arrays are stable. The minimization of the electric energy could only drive the lateral movement of the vortex core to form close pairs. As the field increases, the skyrmion structure starts to appear due to the "melting" of the long vortex line through a Raleigh-Plateau like dynamic process. The skyrmion density increases until it reaches a peak at ~1000 kV/cm. Then, the size of "bubbles" shrinks and eventually burst, leading to the decrease in the skyrmion density at higher field. Consequently, the needle *a* domains inside the bottom layers switch to *c* domains while more and more skyrmions burst to become an *a*/*c* domain structure, following the dimensionality crossover. With a field of ~1500 kV/cm, the skyrmions almost completely disappear, which sets the upper limit for the observation of electric field induced skyrmions.

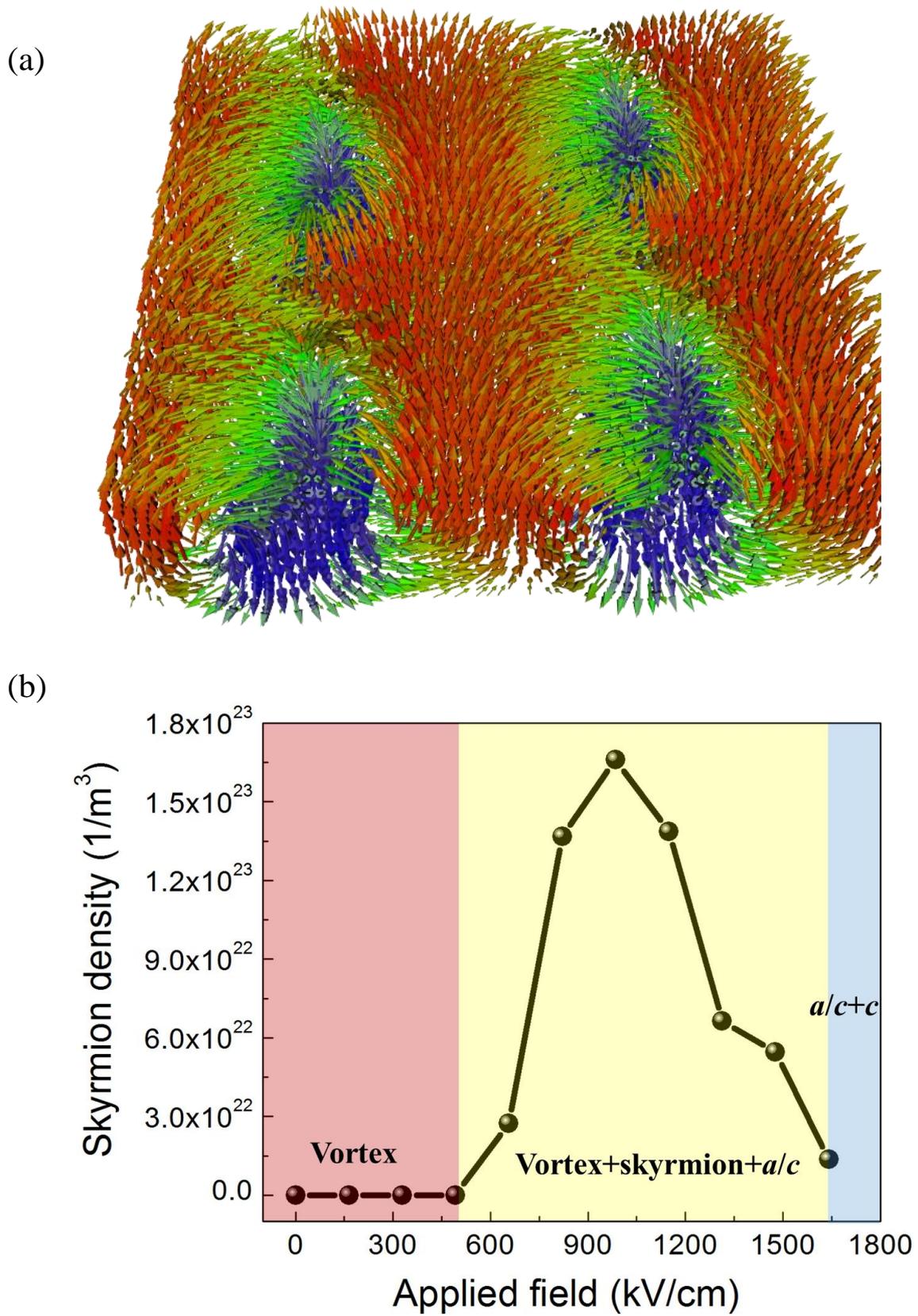

**Fig. 4. 3-D Polar skyrmion structure and stability range.** (a) 3D vector plot of the "bubble" structure, indicating skyrmion-like patterns. The red and blue colors indicate the positive and negative

out-of-plane polarizations, respectively; while the green color shows the in-plane polarizations; (b) Skyrmion density as a function of applied field, showing the stability region of the skyrmion.

To conclude, we have demonstrated the reversible electric switching between two topological structures, the polar vortex lattice and polar skyrmion bubbles, in $(PTO)_{16}/(STO)_{16}$ superlattice on a DSO substrate. At a relatively low bias, the vortex lines moving towards each other to form a close pair so as to reduce the electric energy. Substantially, the two neighboring vortex lines merges, which "melts" to a skyrmion-like "bubble" domain at higher field. This resembles the Pleateau-Raleigh instability in fluid dynamics. The size of the skyrmion-like structure will shrink with the increasing of applied bias which eventually burst at large bias, leading to the formation of ferroelectric/ferroelastic twin domains with distinct 90° domain walls. This switching process involves a dimensionality crossover from 1-D vortex to 2-D domain wall. The electric field phase diagram is constructed, showing the wide stability range of the skyrmion under experimentally accessible field. We hope that this work not only help the understanding of the switching process of the vortex lattice structure in ferroelectric superlattice, but also unveiling the intrinsic tie between the various topological structures (i.e., vortex, skyrmion and meron, ect.) and stimulate further experimental observations and theoretical investigations.

**Acknowledgements**

The work is supported by U.S. Department of Energy, Office of Basic Energy Sciences, Division of Materials Sciences and Engineering under Award FG02-07ER46417 (LQC). Z.J.H acknowledges the support by NSF-MRSEC grant number DMR-1420620 and NSF-MWN grant number DMR-1210588. Z.J.H would like to thank Dr. R. Ramesh and Dr. C. Nelson for helpful discussions.